\documentclass{emulateapj}

\newcommand{\Msun}{M_{\odot}}

\def\gsim{\mathrel{\rlap{\lower 4pt \hbox{\hskip 1pt $\sim$}}\raise 1pt
\hbox {$>$}}}
\def\lsim{\mathrel{\rlap{\lower 4pt \hbox{\hskip 1pt $\sim$}}\raise 1pt
\hbox {$<$}}}

\begin{document}

\title{SN 2006aj Associated with XRF 060218 At Late Phases: \\
Nucleosynthesis-Signature of A Neutron Star-Driven Explosion\altaffilmark{1}}

\author{
K. Maeda\altaffilmark{2}, 
K. Kawabata\altaffilmark{3}, 
M. Tanaka\altaffilmark{4}, 
K. Nomoto\altaffilmark{4,5}, 
N. Tominaga\altaffilmark{4}, 
T. Hattori\altaffilmark{6}, 
T. Minezaki\altaffilmark{7}, 
T. Kuroda\altaffilmark{4}, 
T. Suzuki\altaffilmark{4}, 
J. Deng\altaffilmark{8,4,5}, 
P.A. Mazzali\altaffilmark{4,5,9,10}, 
E. Pian\altaffilmark{10}
}

\altaffiltext{1}{Based on data collected at Subaru Telescope, which is 
operated by the National Astronomical Observatory of Japan.}
\altaffiltext{2}{Department of Earth Science and Astronomy,
Graduate School of Arts and Science, University of Tokyo, 
3-8-1 Komaba, Meguro-ku, Tokyo
153-8902, Japan: maeda@esa.c.u-tokyo.ac.jp}
\altaffiltext{3}{Hiroshima Astrophysical Science Center, 
Hiroshima University, Hiroshima 739-8526, Japan}
\altaffiltext{4}{Department of Astronomy, School of Science, 
University of Tokyo, Bunkyo-ku, Tokyo 113-0033, Japan}
\altaffiltext{5}{Research Center for the Early Universe, School of Science, 
University of Tokyo, Bunkyo-ku, Tokyo 113-0033, Japan}
\altaffiltext{6}{Subaru Telescope, National Astronomical Observatory 
of Japan, Hilo, HI 96720, USA}
\altaffiltext{7}{Institute of Astronomy, School of Science, University of Tokyo, 
2-21-1 Osawa, Mitaka, Tokyo 181-0015, Japan}
\altaffiltext{8}{National Astronomical Observatories, CAS, 
20A Datun Road, Chaoyang District, Beijing 100012, China}
\altaffiltext{9}{Max-Planck-Institut f\"ur Astrophysik, 
Karl-Schwarzschild-Stra{\ss}e 1, 85741 Garching, Germany}
\altaffiltext{10}{National Institute for Astrophysics--OATs, 
Via G.B. Tiepolo 11, 34143 Trieste, Italy}

\begin{abstract}
Optical spectroscopy and photometry of SN 2006aj have been performed 
with the Subaru telescope at $t > 200$ days after GRB060218, 
the X-ray Flash with which it was associated. 
Strong nebular emission-lines with an expansion velocity 
of $v \sim 7,300$ km s$^{-1}$ were detected. 
The peaked but relatively broad [OI]$\lambda\lambda$6300, 6363 
suggests the existence of $\sim 2\Msun$ of materials in which 
$\sim 1.3\Msun$ is oxygen. 
The core might be produced by a mildly asymmetric explosion.  
The spectra are unique among SNe Ic in 
(1) the absence of [CaII]$\lambda\lambda$7291, 7324 emission, and  
(2) a strong emission feature at $\sim 7400$\AA, which requires $\sim 0.05\Msun$ 
of newly-synthesized $^{58}$Ni. 
Such a large amount of stable neutron-rich Ni strongly indicates 
the formation of a neutron star. 
The progenitor and the explosion energy are constrained to 
$18\Msun \lsim M_{\rm ms} \lsim 22\Msun$ and $E \sim (1 - 3) \times 10^{51}$ erg, 
respectively. 
\end{abstract}


\section{INTRODUCTION}

The connection between long soft 
Gamma-Ray Bursts (GRBs) and Type Ic Supernovae (SNe Ic) 
has been observationally established 
(see Woosley \& Bloom 2006 and references therein). 
It has been argued that well examined GRB-associated SNe 
(GRB-SNe) are 
very energetic explosions, called hypernovae, 
resulting from a massive stellar death. 
The kinetic energy ($E_{\rm K}$) is 
$E_{51} \equiv E_{\rm K}/10^{51}$ erg s$^{-1} \gsim 10$, 
and the main-sequence mass of the progenitor 
is $M_{\rm ms} \sim 40\Msun$ (e.g., SN 1998bw: Iwamoto et al. 1998). 

Recently, a new connection between the X-Ray Flash (XRF), 
which is a low energy analog of a GRB, and an SN Ic has been found. 
In association with XRF 060218 (Campana et al. 2006), 
SN Ic 2006aj was discovered 
(Ferrero et al. 2006; Mirabal et al. 2006; Modjaz et al. 2006; 
Pian et al. 2006; Soderberg et al. 2006; Sollerman et al. 2006). 

At early phase, SN 2006aj showed velocities 
intermediate between GRB-SNe and canonical SNe Ic. 
The OI$\lambda$7774 absorption was weak. 
The light curve evolved more quickly than other GRB-SNe. 
Mazzali et al. (2006) concluded that 
SN 2006aj is less energetic ($E_{51} \sim 2$) 
than other GRB-SNe and ejected a smaller amount of $^{56}$Ni 
[$M$($^{56}$Ni) $\sim 0.2\Msun$]. They suggested that 
the main-sequence mass of the progenitor is 
$M_{\rm ms} \sim 20 - 25\Msun$, and hence 
that XRF-associated SN 2006aj was driven by a neutron star (NS) formation. 

In this {\sl Letter}, we present observations of SN Ic 2006aj at $t \gsim 200$ days 
with the Subaru telescope (hereafter $t$ is the rest frame time 
since 2006 February 18). 
We show that the nebular phase data provide 
constraints on nucleosynthesis and the compact remnant. 

\section{Observations and Data Reduction}
We performed spectroscopic and photometric observations of 
SN 2006aj on 2006 September 17 (UT: $t = 204$ days) 
and November 27 ($273$ days) 
with the 8.2 m Subaru telescope equipped with the Faint Object 
Camera and Spectrograph (FOCAS: Kashikawa et al. 2002). 
For spectroscopy, we used $0\farcs 8$ width slit
and the B300 grism, which gave a wavelength coverage
of 4700--9000 \AA\ and a spectral resolution of
$\simeq 10.7$ \AA. 
The exposure times were 2,800 s 
and 5,400 s, respectively. 
For flux calibration, 
GD71 and BD +28$^{\circ}$4211 were observed. 

For photometry, we obtained $B$- (60 s), $V$- (60 s) and $R$-band (60 s) 
images in September 
and $B$- (300 s), $V$- (180 s), and $R$-band (180 s) images in November. 
We obtained images of standard stars around PG0220+132 and 
SA98-634 for photometric calibrations. 
Our photometry is shown in Table 1 and 
the photometry-calibrated spectra in Figure 1.  

The spectra are the sum of the SN and host galaxy spectra (Fig. 1). 
We find broad emission lines with the corresponding expansion 
velocity $v \sim 7,300$ km s$^{-1}$ at both epochs (Fig. 2: see also 
Foley et al. 2006). 
Especially strong is [OI]$\lambda\lambda$6300, 6363. 
So the existence of a SN Ic 
in XRF 060218 is unambiguously confirmed. 

The continuum level is almost identical in both epochs. 
Since no strong emission from an SN is expected 
at $\sim 6100$, $6800$, and $8000$\AA, we regard the flux level 
of the continuum connecting these wavelengths to be the host 
contribution, expressed as 
$-2.0 \times 10^{-17} \times \left(\lambda/6000{\rm \AA}\right) 
+ 4.1 \times 10^{-17}$ erg s$^{-1}$ cm$^{-2}$ \AA$^{-1}$ (Fig. 1). 
The derived host magnitude is fainter than 
that estimated by Sollerman et al. (2006) by about 0.3 mag 
in $V$- and $R$-bands, while the color $V-R$ is consistent. 
The difference probably stems from the systematic errors. 
The host is blue, as expected from its faintness and 
low metallicity (Sollerman et al. 2006).

\section{Analyses of Spectrum Features}
Figure 2 shows that the nebular spectra of SN 2006aj have 
the following two unique features 
compared with other SNe Ic. 
(1) [CaII]$\lambda\lambda$7291, 7324 is absent. 
(2) A strong emission feature is seen at $\sim 7,400$\AA. 
In this {\sl Letter}, we adopt 
$\mu = 35.84$ and $E(B-V) = 0.13$ (Pian et al. 2006). 
Our discussion is based on the September spectrum. 

\subsection{Total Mass at $v \lsim 7,300$ km s$^{-1}$}
The bolometric luminosity ($L_{\rm bol}$) is 
expressed as (e.g., Maeda et al. 2003a),  
\begin{equation}
L_{\rm bol} = (D_{\gamma} + 0.035) L_{\gamma} \ ,
\end{equation} 
at a nebular epoch. Here the $\gamma$-ray energy input rate from $M$($^{56}$Ni) $= 0.2\Msun$ is 
$L_{\gamma} = 4.4 \times 10^{41}$ erg s$^{-1}$ at $t \sim 204$ days. 
$D_{\gamma}$ ($\sim 3\tau_{\gamma}/4$ for a homogeneous nebula, which is a good 
approximation at low velocities) 
is the fraction of the $\gamma$-ray energy deposited within the SN nebula, and 
$0.035$ accounts for the {\it in situ} positron deposition rate. 
The optical depth to the $\gamma$-rays is 
\begin{equation}
\tau_{\gamma} = 0.071 \left(\frac{M_{7300}}{\Msun}\right) 
\left(\frac{v}{7,300 {\rm km s}^{-1}}\right)^{-2} 
\left(\frac{t}{204 {\rm days}}\right)^{-2} \ ,
\end{equation} 
where $M_{7300}$ denotes the total mass ejected with $v \lsim 7,300$ km s$^{-1}$. 
Substituting $L_{\rm bol} \sim 6.2 \times 10^{40}$ erg s$^{-1}$ (taking into account 
the reddening and a typical NIR contribution $\sim 20\%$; Tomita et al. 2006) 
into equations (1) and (2), we obtain $M_{7300} \sim 2\Msun$. 

\subsection{Absence of Ca Lines and Ejecta Structure}
No detection of [CaII] and CaII IR triplet provides the following 
constraints. 
In a core-collapse SN explosion, 
more than $4 \times 10^{-3}\Msun$ of Ca 
is synthesized if $E_{51} \gsim 1$ 
(e.g., Nakamura et al. 2001). 
In an SN Ic, 
the emissivity of [CaII]$\lambda\lambda$7291, 7324 is about three orders of magnitudes larger 
than [OI]$\lambda\lambda$6300, 6363. 
If Ca co-existed with O, Ca$^{+}$ would be 
the predominant ion (e.g., Fransson \& Chevalier 1989). 
Thus, {\it if} Ca ($\gsim 4 \times 10^{-3}\Msun$) 
and O were {\it microscopically} mixed, 
then $M_{\rm O}$ would have been much larger than $20\Msun$ 
to satisfy the observed upper limit for the [CaII] (Fig. 2). 
This is inconsistent with $M_{7300} \sim 2\Msun$ (\S 3.1). 

The separation of Ca and O occurs naturally in ordinary supernova models. 
This is also seen in {\it Spitzer} observations of Cas A (Ennis et al. 2006). 
Nucleosynthesis calculations show that the explosion 
produces chemically distinct regions: 
Fe-rich materials (where $^{56}$Ni is the predominant product), 
Si-rich materials (where most of Ca is produced), 
and O-rich materials (mostly O and some unburned C). 
These regions may well be mixed {\it macroscopically} 
in velocity space (e.g., by jets: Maeda \& Nomoto 2003b; Maeda 2006b), 
but not {\it microscopically}. 
Hereafter, $M_{\rm Fe-rich}$, $M_{\rm Si-rich}$, and $M_{\rm O-rich}$, 
respectively, denote the masses of the Fe-rich, Si-rich, and O-rich materials 
ejected with $v \lsim 7,300$ km s$^{-1}$. 

In this "microscopically separated" model, 
the sum of the [CaII] and CaII luminosities ($L_{\rm CaII}$) is
related to the mass of the Si-rich region rather than 
the mass of Ca itself, i.e., 
\begin{equation}
L_{\rm Ca II} \sim L_{\gamma} X_{\rm Si-rich} D_{\gamma} \alpha_{\rm Si} 
f_{\rm Ca II} \ ,
\end{equation} 
where the fraction of the absorbed $\gamma$-ray energy going into 
the Ca lines would be $f_{\rm CaII} \sim 0.5$, with the remaining fraction 
mostly going into [SiI] NIR lines. 
$X_{\rm Si} \equiv M_{\rm Si-rich}/M_{7300}$ is the mass fraction of 
the Si-rich region. 
A factor, $\alpha_{\rm Si}$, accounts 
for the fact that the $\gamma$-rays 
are scattered more easily within the Fe-rich region 
than in the other regions, because the $\gamma$-rays begin their flight 
in the Fe-rich region. 
At late epochs, the mean free path of the $\gamma$-rays is larger 
than the size of the nebula, thus $\alpha_{\rm Si}$ is an order of unity. 
Taking the upper limit $L_{\rm CaII} 
\lsim 7 \times 10^{39}$ erg s$^{-1}$, 
we obtain 
\begin{equation}
M_{\rm Si-rich} \lsim 0.6\Msun (1/\alpha_{\rm Si}) (0.5/f_{\rm CaII}) \ .
\end{equation} 
The upper limit is even tighter according to 
the spectrum synthesis (\S 3.5), i.e.,   
$M_{\rm Si-rich} \lsim 0.15\Msun$, because other lines 
(e.g., [FeII]) should also have large contribution.

\subsection{Masses of Oxygen and Fe-peak Elements}
From the masses we have obtained as 
$M_{7300} \sim 2\Msun$, $M_{\rm Si-rich} \lsim 0.15\Msun$, 
and $M$($^{56}$Ni) $\sim 0.2\Msun$, we derive 
$M_{\rm O-rich} \sim 1.6\Msun$ and $M_{\rm Fe-rich} \sim 0.3\Msun$. 
We here examine if $M_{\rm O-rich}$ and $M_{\rm Fe-rich}$ 
are consistent with 
the reddening-corrected luminosity of [OI]$\lambda\lambda$6300,6363 ($L_{\rm OI}$) and 
that of [FeII] and [NiII] ($L_{\rm Fe-peak}$: 
integrated in 4700 - 5500\AA\ and 7000 - 7600\AA). 

Following the argument similar to the one used in deriving equation (4), we obtain 
\begin{equation}
M_{\rm O-rich} \sim  1.6\Msun (0.7/\alpha_{\rm O}) (0.7/f_{\rm OI}) \ ,
\end{equation} 
with $L_{\rm OI} \sim 1.7 \times 10^{40}$ erg s$^{-1}$. 
Hereafter, 
$\alpha_{\rm i}$ and $X_{\rm i}$ (where i = Fe, Si and O) are used for the Fe-, Si-, and 
O-rich regions in the same way as $\alpha_{\rm Si}$ and $X_{\rm Si-rich}$. 
$f_{\rm OI}$ is the fraction of the absorbed $\gamma$-ray energy going into 
[OI]$\lambda\lambda$6300,6363 (with remaining fraction going into 
[CI]$\lambda$8727 and OI$\lambda$7774). 
Equation (5) shows that $\alpha_{\rm O} \sim 0.7$ is necessary to reproduce $L_{\rm OI}$.  
To meet the trivial relations $\sum_{\rm i} X_{\rm i} 
\alpha_{\rm i} =1$ and $\sum_{\rm i} X_{\rm i} = 1$, 
$\alpha_{\rm Fe} \sim 2.6$ is required. 
With this value, the spectrum synthesis (\S 3.5) well reproduces 
$L_{\rm Fe-peak} \sim 2 \times 10^{40}$ ergs s$^{-1}$. 

These values for $\alpha_{\rm i}$ are physically reasonable. 
For an SN nebula having a constant density 
with $M_{\rm Fe-rich} = 0.3\Msun$, $M_{\rm Si-rich} = 0.1\Msun$, and 
$M_{\rm O-rich} = 1.6\Msun$, we derive 
$\alpha_{\rm Fe-rich} \sim 2.9$, $\alpha_{\rm Si-rich} \sim 1.8$, and 
$\alpha_{\rm O-rich} \sim 0.6$ by our 1D $\gamma$-ray transport calculation. 
These are obtained for a concentric distribution (with the Fe region 
in the center), but it gives an approximately correct estimate even 
for a jet/blob-like distribution of the Fe-rich materials. 
 
\subsection{[NiII]$\lambda$7380 and stable $^{58}$Ni}
The strong emission at $\sim 7400$\AA\ is not common for SNe Ib/c, 
although in other SNe Ib/c the feature may be hidden by 
the [CaII]. 
The contribution from [FeII]$\lambda$7388 and [FeII]$\lambda$7452 (a$^{4}$F-a$^{2}$G) 
should be small. 
The radiative transition probabilities of the [FeII] are smaller than
those of [FeII]$\lambda$7172 and [FeII]$\lambda$7155 that arise from the same 
transitions, and the luminosity at $7100 - 7200$\AA\ 
is only one-third of the luminosity of the $7400$\AA\ feature. 

We suggest that the $7400$\AA\ feature is [NiII]$\lambda$7380 (Fig. 2), 
as also seen in the Crab Nebula (MacAlpine et al. 2007). 
From the spectrum synthesis (\S 3.5), 
we find that this feature is reproduced by 
[NiII]$\lambda$7380 if the mass of Ni is $\sim 0.05\Msun$. 
Since virtually all $^{56}$Ni already decayed at $t \gsim 200$ days, 
this must be stable $^{58}$Ni.

\subsection{Spectrum Synthesis}
To confirm the above results, we perform nebular spectrum synthesis 
calculations. 
We set the masses at $v \leq 7,300$ km s$^{-1}$ as follows: 
$M_{\rm O-rich} = 1.6\Msun$ (where $M_{\rm O} = 1.3\Msun$ and $M_{\rm C} = 0.3\Msun$), 
$M_{\rm Si-rich} = 0.1\Msun$ (where $M_{\rm Si} = 0.08\Msun$, $M_{\rm S} = 0.01\Msun$, 
$M_{\rm Ca} = 4 \times 10^{-3}\Msun$, and $M_{\rm Fe} = 4 \times 10^{-3}\Msun$), 
and $M_{\rm Fe-rich} = 0.3\Msun$ 
[where $M$($^{56}$Ni) $= 0.2\Msun$ and $M$($^{58}$Ni) $= 0.06\Msun$].  
We set $\alpha_{\rm Fe} = 2.6$, $\alpha_{\rm Ca} = 1.0$, and $\alpha_{\rm O} = 0.7$ 
for the $\gamma$-ray deposition (\S 3.3). 

A one-zone spectrum synthesis is applied to the O-rich, Si-rich, 
and Fe-rich materials. 
A Monte-Carlo $\gamma$-ray transport calculation is performed for a uniform 
nebula with $M_{7300} = 2\Msun$, $M$($^{56}$Ni) $=0.2\Msun$. 
The deposited luminosity is apportioned to each region 
according to $X_{\rm i} \alpha_{\rm i}$ (equation 3). Positrons deposit their energy 
only to the Fe-rich materials. 

With the deposition luminosity, non-LTE rate equations coupled with the 
ionization-recombination equilibrium are solved for each region 
(Mazzali et al. 2001). 
The model spectrum fits the observed one fairly well (Fig. 3). 
The feature at $\sim 7,400$A is well explained by [NiII]$\lambda$7380 (\S 3.3) 
from $\sim 0.05\Msun$ of stable Ni.

\section{Properties of the Progenitor and its Explosion}
We have derived the masses as follows (\S 3): 
$M_{7300} \sim 2\Msun$, 
$M_{\rm O-rich} \sim 1.6\Msun$ 
(i.e., $M_{\rm O} \sim 1.3\Msun$ if $80\%$ is oxygen), 
$M_{\rm Si-rich} \lsim 0.15\Msun$, and $M_{\rm Fe-rich} \sim 0.3\Msun$ 
(where $M$($^{56}$Ni) $\sim 0.2\Msun$ and $M$($^{58}$Ni) $\sim 0.05\Msun$). 

{\bf Progenitor Mass:} 
The masses of 
$M_{\rm O} \sim 1.3\Msun$ and $M_{7300} \sim 2\Msun$ provide 
strong constraints on $M_{\rm ms}$ as follows. 
Theoretical models show $M_{\rm O} < 1.3\Msun$ 
in the whole ejecta if $M_{\rm ms} \lsim 18\Msun$ (Nomoto \& Hashimoto 1988). 
This leads to the lower limit of $M_{\rm ms} \gsim 18\Msun$. 

For given $E_{51}$, larger $M_{\rm ms}$ results in 
larger $M_{7300}$. 
Our spherical calculations show 
$M_{7300} > 2\Msun$ if $M_{\rm ms} \gsim 22\Msun$ and 
$E_{51} = 2$. 
Since explosion asymmetry 
only increases the mass at the low velocity 
(Maeda et al. 2002; Maeda \& Nomoto 2003b), 
the result of the spherical 
calculations can be used to place the upper limit to $M_{\rm ms}$. 
We thus obtain $M_{\rm ms} \lsim 22\Msun$. 
Therefore, we narrow down the progenitor mass range as  
$18\Msun \lsim M_{\rm ms} \lsim 22\Msun$. 

The total oxygen mass is $\sim 2.3\Msun$ in the ejecta, 
since the oxygen mass at $v \gsim 7,300$km s$^{-1}$ 
is $\sim 1\Msun$ (Mazzali et al. 2006). 
This is consistent with $M_{\rm ms} \sim 22\Msun$. 

{\bf Explosion Energy:}
$M_{\rm Si-rich} \lsim 0.15\Msun$ (\S 3.2) gives information on 
$E_{51}$. 
The mass of the Si-rich region is larger for larger $M_{\rm ms}$ 
(because of the higher progenitor density) 
and for larger $E_{51}$ (because of the more extended burning). 
According to our explosion models (e.g., Nakamura et al. 2001), 
we obtain the constraints of $M_{\rm ms} \lsim 25\Msun$ and 
$E_{51} \lsim 3$. 

{\bf NS Formation:}
$M$($^{58}$Ni) $\sim 0.05\Msun$ provides a 
strong indication of the NS formation. 
The large amount of $^{58}$Ni indicates that neutron-rich materials are ejected from the vicinity of 
the central remnant. 
Our models for $M_{\rm ms} = 20\Msun$ and $E_{51} \sim 2.5$ 
shows that $M$($^{56}$Ni) $\sim 0.2\Msun$ and $M$($^{58}$Ni) $\sim 0.04\Msun$ 
if the remnant's mass is $M_{\rm cut} \sim 1.4\Msun$.  
If $M_{\rm cut} \sim 1.5\Msun$, then $M$($^{56}$Ni) $\sim 0.16\Msun$ and 
$M$($^{58}$Ni) $\sim 9 \times 10^{-4}\Msun$. 
For $M_{\rm cut} \sim 1.6\Msun$, $M$($^{56}$Ni) $\sim 0.07\Msun$ and 
$M$($^{58}$Ni) $\sim 3 \times 10^{-4}\Msun$. 
The NS formation is also consistent with $M$($^{56}$Ni) $\sim 0.2\Msun$.

\section{Discussion}

\subsection{Signatures of an Asymmetric Explosion}
In GRB-SN 1998bw, the observed mass at low velocities exceeds 
that predicted by the spherical explosion model, 
which suggests that the explosion is aspherical (Mazzali et al. 2001; 
Maeda et al. 2003a, 2006ac). On the other hand, the existence of 
the asymmetry is not clear in SN 2006aj. 
$M_{7300} \sim 2\Msun$ (\S 3) is consistent with 
the spherical explosion model for a CO 
star with $M_{\rm ms} \sim 22\Msun$ and $E_{51} \sim 2$. 

We note, however, that 
the spherical model has a low density hole at $v \lsim 2,000 - 2,500$ km s$^{-1}$.   
The mildly-peaked [OI]$\lambda\lambda$6300,6363 profile shows 
an enhancement of the material density at $v \lsim 3,000$ km s$^{-1}$.  
This indicates an asymmetric explosion, as also suggested 
by polarizations (Gorosabel, et al. 2006). 
The asymmetry might be produced either by a jet-like explosion 
or Rayleigh-Taylor instability at the Si/CO interface 
(Kifonidis et al. 2003). 
If this is a signature of a jet-like aspherical SN explosion, 
the jet would be wider than in SN 1998bw 
(intrinsically or due to stronger lateral expansion; 
Maeda \& Nomoto 2003b), 
since the signature is seen only in the innermost part 
(see also Mazzali et al. 2007). 

\subsection{Relations to other SNe Ib/c and SN 1998bw}
The absence of [CaII]$\lambda\lambda$7291, 7324 indicates that the 
chemically distinct regions 
are not microscopically mixed to one another (\S 3.2). This is probably the case 
in SNe Ib/c in general: otherwise the [CaII] would be much stronger than 
[OI]$\lambda\lambda$6300, 6363 in most SNe Ib/c, as is contrary to observations 
(e.g., Matheson et al. 2001). 

Why the [CaII] is absent only in SN 2006aj may be answered from a tendency of 
$E_{51}$ as a function of $M_{\rm ms}$.  
For progenitors with $M_{\rm ms} \lsim 20\Msun$, the [CaII] can be strong 
relative to the [OI], because $M_{\rm O-rich}$ is smaller than in SN 2006aj 
despite similar $M_{\rm Si-rich}$ (since $E_{51} \sim 1 -2$). 
For hypernovae with $M_{\rm ms} \sim 40\Msun$ (e.g., SN 1998bw), 
the [CaII] can be strong since the large $E_{51} \gsim 10$ yields large $M_{\rm Si-rich}$. 
Thus, presence of the [CaII] in the majority of normal SNe Ib/c suggests that  
such normal SNe Ib/c may have originated in relatively low-mass stars 
with $M_{\rm ms} \lsim 20\Msun$. 
This implies that they evolve through binary paths, since single stars 
with this mass range are expected to end as SNe II (Hirschi et al. 2005).

\acknowledgements

K.M. and N.T. are JSPS Research Fellows. 
J.D. is supported by NSFC grant 10673014.

\clearpage
\begin{deluxetable}{cccc}
 \tabletypesize{\scriptsize}
 \tablecaption{Photometry
 \label{tab:phot}}
 \tablewidth{0pt}
 \tablehead{
   \colhead{Date/Component}
 & \colhead{$B$}
 & \colhead{$V$}
 & \colhead{$R$}
}
\startdata
September 17\tablenotemark{a} & $20.64 \pm 0.1$ & $20.23 \pm 0.1$ & $19.82 \pm 0.1$\\
November 27\tablenotemark{a} & $20.71 \pm 0.1$ & $20.34 \pm 0.1$ & $19.86 \pm 0.1$\\
Host (Subaru)\tablenotemark{b} & \nodata        & $20.53$          & $20.16$\\
SN on September\tablenotemark{c} & \nodata   & $22.45$          & $21.74$\\
SN on November\tablenotemark{c} & \nodata               & $22.94$          & $22.21$\\
Host (Sollerman)\tablenotemark{d} & $20.46 \pm 0.07$ & $20.19 \pm 0.04$ & $19.86 \pm 0.03$\\
\enddata
\tablenotetext{a}{Using the FOCAS filter system, which is similar to 
the Johnson-Cousins system (Kashikawa et al. 2002).}
\tablenotetext{b}{See \S 2 and Figure 1 for the assumed host spectrum. 
The photometry is given for the Johnson-Cousins system.}
\tablenotetext{c}{The intrinsic SN magnitude obtained after subtracting the host 
contribution and removing narrow emission lines. In the Johnson-Cousins system. }
\tablenotetext{d}{The host magnitude estimated by Sollerman et al. (2006)}
\end{deluxetable}

\clearpage
\begin{figure}
\begin{center}
	\begin{minipage}{0.8\textwidth}
		\epsscale{1.0}
		\plotone{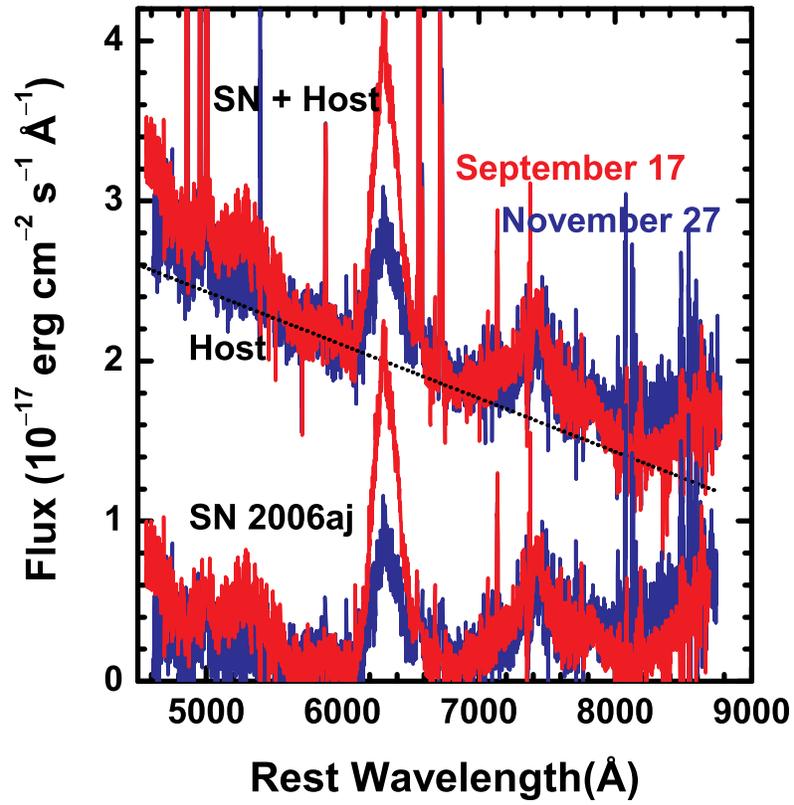}
	\end{minipage}
\end{center}
\caption[] 
{Reduced spectra of SN 2006aj on 2006 September 17 (red), 
and on 2006 November 27 (blue). The flux is calibrated with 
the photometry. 
The redshift of the host ($z = 0.0335$) is corrected for. 
The assumed host galaxy contribution (\S 2) is shown by 
the black dotted line. 
The intrinsic SN spectra, after subtracting the host contribution and 
removing major narrow emission lines, are also shown (bottom). 
\label{fig1}}
\end{figure}

\clearpage
\begin{figure}
\begin{center}
\begin{minipage}{0.6\textwidth}
		\epsscale{1.0}
		\plotone{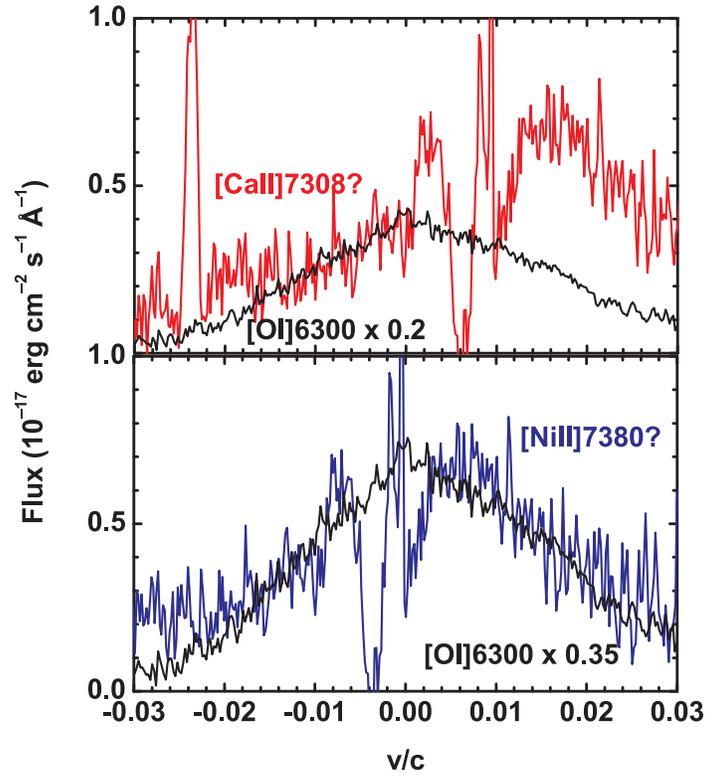}
	\end{minipage}
\end{center}
\caption[] 
{Comparison of emission profiles. 
(a) Comparison of [OI]$\lambda\lambda$6300, 6363 (black)
with [CaII]$\lambda\lambda$7291, 7324 (red). The center of 
the wavelength is assumed to be $6300$ and $7308$\AA\ for 
the [OI] and [CaII], respectively. 
(b) Comparison of the [OI] (black) with the emission feature 
centered at $7380$\AA (blue). 
We suggest that the feature is [NiII]$\lambda$7380. 
Note that an atmospheric absorption at $\sim 7,400$\AA\ 
is not removed. 
\label{fig2}}
\end{figure}

\clearpage
\begin{figure}
\begin{center}
	\begin{minipage}{0.8\textwidth}
		\epsscale{1.0}
		\plotone{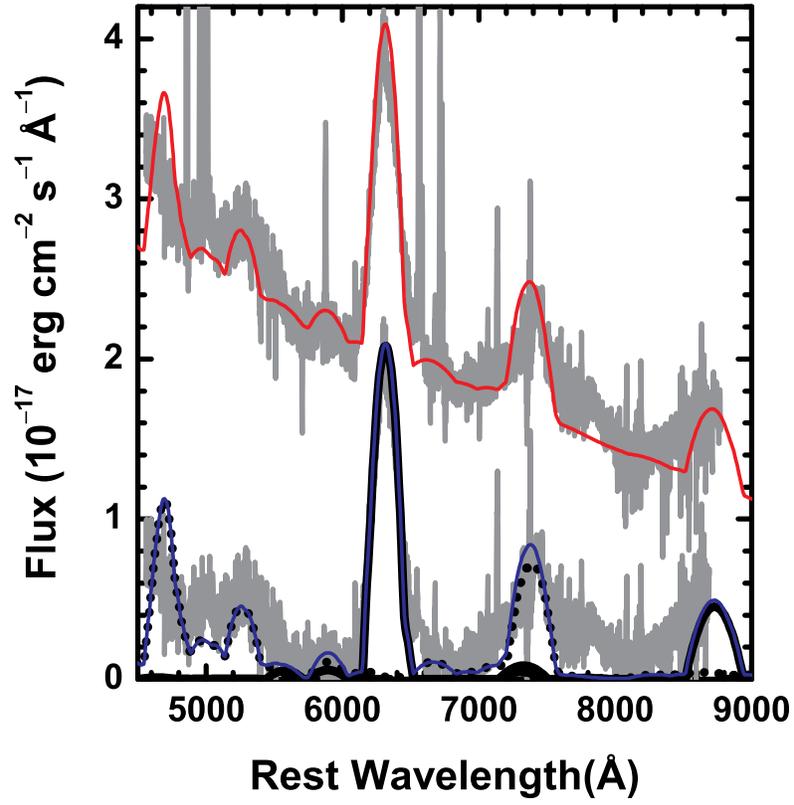}
	\end{minipage}
\end{center}
\caption[] 
{Synthetic spectrum as compared with the September spectrum. 
The model spectrum, reddened with $E(B-V) = 0.13$, 
is compared with the host-subtracted SN spectrum at the SN rest frame 
(blue). For presentation, the fit to the original spectrum 
is also shown (red), in which the host spectrum (Fig. 1) is added 
to the model spectrum. 
For the model spectrum, the individual contributions from the O-rich (black solid), 
Si-rich (dashed), and Fe-rich (dotted) regions are shown.  
\label{fig3}}
\end{figure}

\end{document}